\pdfoutput=1
\documentclass[showpacs,superscriptaddress,floatfix,aps,prl,notitlepage,reprint]{revtex4-2}
\usepackage{graphicx,amsfonts,amssymb,amsmath,hyperref,hypcap,enumerate}
\usepackage{braket}
\usepackage{amsthm}
\usepackage{multirow}
\usepackage{dcolumn}
\usepackage{bm}
\usepackage{epstopdf}
\usepackage{mathtools}

\usepackage{color}
\usepackage{subfigure}
\usepackage{diagbox}
\usepackage{longtable}
\usepackage{makecell}
\usepackage{bbold}
\usepackage{algorithm}
\usepackage{algpseudocode}
\usepackage{svg}
\usepackage{tikz}
\usetikzlibrary{decorations.pathreplacing,calc}

\DeclareMathAlphabet{\mathitb}{OT1}{cmr}{bx}{sl}
\usepackage[utf8]{inputenc}
\usepackage{enumitem}

\newcommand{\mO}{\mathcal{O}}

\newcommand{\Tr}{\mathrm{Tr}}
\newcommand{\ii}{\mathrm{i}}
\newcommand{\mW}{\mathbb{W}}

\newcommand{\bs}{\mathbf s}
\newcommand{\ts}{\tilde{\mathbf s}}
\newcommand{\ann}[1]{\overline{#1}^{\mathrm{A}}}
\newcommand{\dket}[1]{\mathinner{\lvert #1 \rangle\!\rangle}}
\newcommand{\dbra}[1]{\mathinner{\langle\!\langle #1 \rvert}}
\newcommand{\dbraket}[2]{\mathinner{\langle\!\langle #1 \mid #2 \rangle\!\rangle}}
\newcommand{\drho}{\dket{\overline{\rho_t^{\otimes 4}}}}

\definecolor{mycoral}{RGB}{224,122,95}
\definecolor{myblue}{RGB}{66,135,245}
\definecolor{myyellow}{RGB}{240,188,66}
\definecolor{mybond}{RGB}{0,0,0}
\newcommand{\twositegate}{
\mathord{\vcenter{\hbox{
\begin{tikzpicture}[scale=1]
  \path[use as bounding box] (-1.9,0.20) rectangle (-1.1,-0.44);
  \draw[thick] (-1.75,0.08) -- (-1.75,0.20);
  \draw[thick] (-1.25,0.08) -- (-1.25,0.20);
  \draw[thick] (-1.75,-0.32) -- (-1.75,-0.44);
  \draw[thick] (-1.25,-0.32) -- (-1.25,-0.44);
  \draw[thick, fill=myblue, rounded corners=2pt] (-1.9,0.08) rectangle (-1.1,-0.32);
\end{tikzpicture}
}}}
}
\newcommand{\init}{
\mathord{\vcenter{\hbox{
\begin{tikzpicture}[scale=1]
  \path[use as bounding box] (-0.22,0.22) rectangle (0.22,-0.22);
  \draw[thick, fill=myyellow, rounded corners=1pt] (-0.22,0.22) rectangle (0.22,-0.22);
  \draw[thick] (0,0.22) -- (0,0.40);
\end{tikzpicture}
}}}
}
\definecolor{mygreen}{RGB}{104,176,99}
\newcommand{\QQ}{
\mathord{\vcenter{\hbox{
\begin{tikzpicture}[scale=1]
  \path[use as bounding box] (-0.22,0.22) rectangle (0.22,-0.22);
  \draw[thick, fill=mygreen, rounded corners=1pt] (-0.22,0.22) rectangle (0.22,-0.22);
  \draw[thick] (0,-0.22) -- (0,-0.40);
\end{tikzpicture}
}}} 
}

\newcommand{\drhoImpsDiagram}{%
\mathord{\vcenter{\hbox{%
\begin{tikzpicture}[
  baseline=(current bounding box.center),
  x=1cm,
  y=1cm,
  line width=0.85pt,
  line cap=round,
  line join=round
]
  \def\dx{0.64}%
  \def\gammaHalf{0.22}%
  \def\lambdaHalfX{0.150}%
  \def\lambdaHalfY{0.165}%
  \def\legLen{0.66}%
  \def\lambdaLabelY{-0.64}%
  \def\gammaLabelY{-0.64}%
  \def\dotsLeftOffset{0.22}%
  \def\dotsRightOffset{0.62}%
  \def\dotsSep{0.085}%
  \def\dotsRadius{0.018}%
  \def\leftBondOverhang{-0.1}%
  \def\rightBondOverhang{0.22}%
  \def\bboxY{1.02}%
  \foreach \n in {0,...,4} {
    \coordinate (x\n) at ({\n*\dx},0);
  }
  \coordinate (dotsL) at (-\dotsLeftOffset,0);
  \coordinate (dotsR) at ({4*\dx+\dotsRightOffset},0);
  \path[use as bounding box]
    (-0.50,-\bboxY) rectangle ({4*\dx+1.06},\bboxY);
  \foreach \dotSign in {-1,0,1} {
    \pgfmathsetmacro{\dotShift}{\dotSign*\dotsSep}%
    \fill ($(dotsL)+(\dotShift,0)$) circle[radius=\dotsRadius cm];
    \fill ($(dotsR)+(\dotShift,0)$) circle[radius=\dotsRadius cm];
  }
  \draw[draw=mybond] (-\leftBondOverhang,0) -- ({4*\dx+\rightBondOverhang},0);
  \foreach \x/\lab in {
    x2/{\Lambda^{[0]}},
    x4/{\Lambda^{[1]}}
  } {
    \draw[fill=white,draw=mybond]
      ($(\x)+(0,\lambdaHalfY)$) --
      ($(\x)+(\lambdaHalfX,0)$) --
      ($(\x)+(0,-\lambdaHalfY)$) --
      ($(\x)+(-\lambdaHalfX,0)$) -- cycle;
    \node at ($(\x)+(0,\lambdaLabelY)$) {$\lab$};
  }
  \foreach \x/\lab in {
    x1/{\Gamma^{[0]}},
    x3/{\Gamma^{[1]}}
  } {
    \draw[fill=mycoral,draw=mybond]
      ($(\x)+(-\gammaHalf,-\gammaHalf)$) rectangle
      ($(\x)+(\gammaHalf,\gammaHalf)$);
    \draw[draw=mybond] ($(\x)+(0,\gammaHalf)$) -- ($(\x)+(0,\legLen)$);
    \node at ($(\x)+(0,\gammaLabelY)$) {$\lab$};
  }
\end{tikzpicture}%
}}}%
}

\newcommand{\drhoNetworkEquation}{%
\drho\;\equiv\;\drhoImpsDiagram
}

\usepackage{etoolbox}
\makeatletter
\patchcmd{\frontmatter@abstract@produce}
  {\vskip200\p@\@plus1fil
   \penalty-200\relax
   \vskip-200\p@\@plus-1fil}
  {}
  {}
  {}
\makeatother

\begin{document}

\title{Diffusive Dynamics of Nonstabilizerness}

\author{Zhenyu Xiao}
\email{zyxiao@princeton.edu}
\affiliation{Princeton Quantum Initiative, Princeton University, Princeton, New Jersey 08544, USA}

\author{Shinsei Ryu}
\affiliation{Department of Physics, Princeton University, Princeton, New Jersey 08544, USA}

\date{\today}

\begin{abstract}
Symmetries shape the quantum-information dynamics of many-body systems, but their effect on nonstabilizerness, the resource complementary to entanglement, is less understood.
We compute the stabilizer R\'enyi entropy, a measure of nonstabilizerness, in $\mathrm{U}(1)$-symmetric one-dimensional random circuits. 
The disorder-averaged dynamics is captured by a four-replica tensor network, which we evaluate by $S_4$-adapted infinite time-evolving block decimation (iTEBD) directly in the thermodynamic limit.
Together with a hydrodynamic argument, our results identify a diffusive universality class for the late-time approach of nonstabilizerness to its random-state value, with the stabilizer R\'enyi entropy gap closing as $1/t$.
The same scaling is verified in an energy-conserving nonintegrable Ising chain.
More broadly, our framework provides a hydrodynamic perspective on nonstabilizerness generation and offers insight into the design of approximate Haar-random states in Hamiltonian dynamics.
\end{abstract}

\maketitle

{\it Introduction.---}
Achieving quantum advantage requires multiple quantum resources~\cite{shor1997,childs2010,chitambar2019}.
A prime example is entanglement, which encodes nonclassical global correlations and is a key ingredient for many quantum algorithms~\cite{amico2008,eisert2010,nielsen2010}.
Entanglement alone, however, is not enough: Clifford circuits can generate extensive entanglement on stabilizer states while remaining efficiently classically simulable~\cite{gottesman1998,aaronson2004}.
This motivates the notion of nonstabilizerness, or quantum magic~\cite{bravyi2005,howard2014,veitch2014,campbell2010,wang2020a,seddon2021}, which measures how far a state departs from the set of stabilizer states.
How these resources are generated and redistributed under quantum dynamics is therefore a central question for quantum computation and engineering, and provides a quantum-information perspective on many-body physics~\cite{rattacaso2023,zhang2024,montanaLopez2024,turkeshi2025,tirrito2025a,odavic2025,bejan2024,bejan2025,maity2025,dowling2025a,aditya2025,fux2024a,tarabunga2025,scocco2025,niroula2024,oliviero2022,tarabunga2024b,white2021a,qian2025a,frau2025,hoshino2025a,leone2021a,jasser2025,zhang2025f,bera2025a,korbany2025a,santra2025,haug2025,odavic2023,passarelli2025,hou2025,sommers2024}.

Symmetries and conservation laws provide a broad organizing principle for many-body quantum dynamics~\cite{dalessio2016,nandkishore2015}.
When the symmetry yields a conserved charge, the associated slow modes control the long-time, long-distance structure of relaxation, with consequences for operator spreading and out-of-time-ordered correlators~\cite{vonkeyserlingk2018,khemani2018,gopalakrishnan2018,jacoby2025,rakovszky2018,nahum2018a}.
They also reshape more global many-body diagnostics: in random circuits with a conserved charge, higher R\'enyi entanglement entropies grow sub-ballistically as $\sqrt{t}$~\cite{rakovszky2019,zhou2020b}, related hydrodynamic constraints appear in charge-transfer statistics and spectral form factors~\cite{mcculloch2023,chan2018,moudgalya2021,shivam2023}, and Mpemba-type effects emerge in resource relaxation~\cite{aditya2025a,xiao2026}.
Slow relaxation has also been reported for participation entropy in symmetry-constrained dynamics~\cite{aditya2026coherence}.

The analogous question for nonstabilizerness --- how symmetry shapes its dynamics --- is essential to a complete understanding of symmetry-resolved many-body physics and its dynamical universality classes~\cite{liu2024,liu2025d,smith2025,sticlet2025,falcao2025b,wang2025,iannotti2026}.
It is also of practical importance: realistic routes to magic generation in quantum computing and simulation are typically structured rather than fully Haar-random~\cite{magni2025,szombathy2025a,szombathy2025b}.
In fermionic quantum computation, trial-state circuits can be designed to remain within fixed particle-number sectors~\cite{gard2020,bravyi2002,collura2025b}, and quantum simulations of gauge theories operate within symmetry-constrained Hilbert spaces~\cite{martinez2016,rajput2023,tarabunga2023manybody,falcao2025a}.

Despite recent progress~\cite{smith2025,sticlet2025,falcao2025b,wang2025,tarabunga2023manybody,falcao2025a,iannotti2026}, how symmetry shapes magic dynamics has remained largely open, in part because of its technical complexity.
Many measures of nonstabilizerness are defined through optimization over stabilizer decompositions and are therefore notoriously hard to compute~\cite{howard2014,beverland2020,liu2022,heinrich2019}.
Recently proposed Pauli-string-based measures are more tractable~\cite{leone2022,haug2023b,haug2023c,beverland2020,jiang2023d,turkeshi2025a,haug2024,ding2025,liu2025c,tarabunga2025a,paviglianiti2025,hinsche2025}, but their cost still scales exponentially in $N$.
Exploring dynamical universality, which requires both late times and large system sizes, is even more challenging: direct matrix-product-state simulations of the time-evolved wavefunction become inefficient once its entanglement reaches the volume law~\cite{lami2024,chen2024a,haug2023a,tarabunga2024a,frau2024b,lami2023,tarabunga2025a}.
This motivates a complementary way to track the dynamics.

In this Letter, we study the dynamics of the stabilizer R\'enyi entropy, a tractable measure of nonstabilizerness, in one-dimensional $\mathrm{U}(1)$-symmetric circuits.
To access the late-time behavior at large system sizes, we derive a replica tensor-network representation of the disorder-averaged dynamics.
The ensemble average restores translation invariance, so we evaluate it directly in the thermodynamic limit by infinite time-evolving block decimation (iTEBD)~\cite{vidal2003,vidal2007,orus2008}, exploiting the $S_4$ replica symmetry for numerical efficiency.
We find that, in the thermodynamic limit, the stabilizer R\'enyi entropy density saturates at late times to its Haar-random value, and the gap to this value closes algebraically as $\Delta m_2(t)\propto t^{-1}$. We trace this scaling to the diffusive relaxation of the conserved charge.
Direct simulations at finite $N$ further reveal that this $t^{-1}$ behavior holds within a diffusive window $\tau\ll t\ll t_D\sim N^2/D$, with $D$ the charge diffusion constant, before crossing over to an exponential tail; higher-order R\'enyi gaps follow the hierarchy $\Delta M_q\propto t^{-q/2}$ in the same window.
The same scaling appears in an energy-conserving nonintegrable Ising chain, identifying a single diffusive universality class for the late-time approach of nonstabilizerness to its random-state plateau.
Our work establishes hydrodynamics as a universal organizing principle for late-time magic dynamics in symmetry-constrained chaotic systems, and provides a tensor-network framework for accessing nonstabilizerness directly in the thermodynamic limit.

\begin{figure*}[t]
  \centering
  \includegraphics[width=0.98\textwidth]{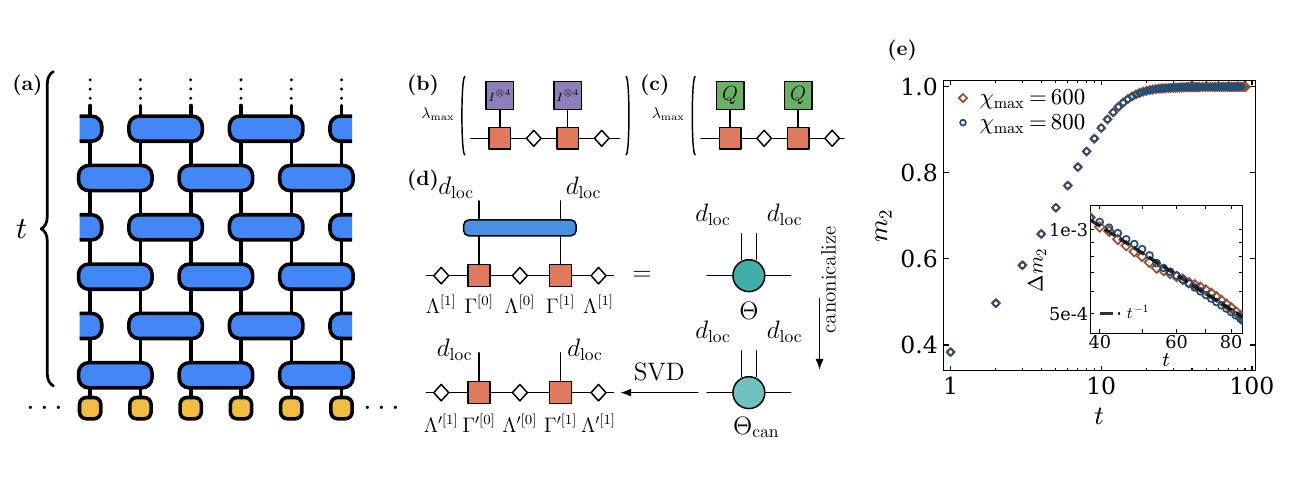}
  \caption{
  (a) Disorder-averaged four-replica transfer network for $\drho$. Blue tensor: Haar-averaged $\mathrm{U}(1)$-symmetric gate $\mW$ [Eq.~\eqref{eq:W4-main}]; yellow tensor: averaged initial state $\dket{\overline{\rho_0^{\otimes 4}}}$ [Eq.~\eqref{eq:rho0}].
  (b,c) Contractions with $I^{\otimes 4}$ and $Q$ defining two-site transfer matrices with dominant eigenvalues $\lambda_{\rm max;tr}$ and $\lambda_{\rm max;Q}$, whose ratio determines the annealed entropy density.
  (d) Two-site update of the infinite matrix product state: apply $\mW$, restore canonical form, and truncate by singular value decomposition within the $S_4$ blocks.
  (e) Annealed stabilizer R\'enyi entropy density $\ann{m_2}$ versus time. Inset: late-time behavior on a log-log scale.
  }
  \label{fig:main-schematic}
\end{figure*}

{\it Nonstabilizerness in $\mathrm{U}(1)$-symmetric dynamics.}---
We study a one-dimensional $\mathrm{U}(1)$-symmetric random circuit on $N$ qubits:
\begin{equation}
  \ket{\psi_t} =\prod_{i=1}^{t} \Big( \prod_{j} U_{2j,2j+1}(i) \prod_{j} U_{2j-1,2j}(i) \Big) \ket{\psi_0},
\end{equation}
where each two-qubit gate $U_{j,j+1}(i)$ is independently drawn from the Haar measure on two-qubit unitaries satisfying $[U_{j,j+1},Z_j+Z_{j+1}]=0$.
In the computational basis $(00,01,10,11)$, this is a block-diagonal $\mathrm{U}(1)\oplus \mathrm{U}(2)\oplus \mathrm{U}(1)$ gate, with the three blocks sampled from independent Haar measures.
We choose a random product initial state $\ket{\psi_0}=\bigotimes_i\ket{\phi_i}$ with each $\ket{\phi_i}$ Haar-random on a single qubit.
The dynamics conserves the total charge $\sum_i Z_i$.

The nonstabilizerness generated along this evolution is quantified by the stabilizer R\'enyi entropy~\cite{leone2022}.
Expanding the pure-state density matrix in the Pauli basis $P\in\{I,X,Y,Z\}^{\otimes N}$ as $\ket{\psi}\!\bra{\psi}=d^{-1}\sum_P c_P P$, with $c_P\equiv\langle\psi|P|\psi\rangle$ and $d=2^N$, the purity condition $\Tr[(\ket{\psi}\!\bra{\psi})^2]=1$ gives $\sum_P c_P^2/d=1$.
Thus, $\{c_P^2/d\}$ is a probability distribution on Pauli strings. The corresponding stabilizer R\'enyi entropy is
\begin{equation}
M_\alpha(|\psi\rangle):=\frac{1}{1-\alpha}\log_2\zeta_\alpha-N,\qquad
\zeta_\alpha\equiv\sum_P\frac{c_P^{2\alpha}}{d^\alpha}.
\label{eq:stabilizer-renyi-entropy-def}
\end{equation}
The shift $-N$ is chosen so that $M_\alpha\ge 0$, with equality iff $|\psi\rangle$ is a stabilizer state. $M_\alpha$ is a stabilizer monotone for pure states when $\alpha\ge 2$~\cite{leone2024}.
We consider $\alpha=2$, and the stabilizer purity admits the four-replica form
\begin{equation} \label{eq:zeta}
\zeta_{2} = \Tr\!\left[ (\ket{\psi}\!\bra{\psi})^{\otimes 4}\, Q\right],\qquad
Q \equiv \frac{1}{d^2}\sum_{P} P^{\otimes 4}.
\end{equation}
With $\rho_t=\ket{\psi_t}\!\bra{\psi_t}$, the ensemble-averaged stabilizer purity $\overline{\zeta_2(t)}=\Tr(Q\,\overline{\rho_t^{\otimes 4}})$ determines the annealed entropy $\ann{M_2}(t)=-\log_2\overline{\zeta_2(t)}-N$.

{\it iTEBD for the replica tensor network.}---
We turn $\rho_t^{\otimes 4}$ into a vector $\dket{\rho_t^{\otimes 4}}$ in a doubled Hilbert space via the state-operator mapping; unitary evolution acts as $U\otimes U^*$ and the trace becomes an inner product, $\Tr(Q\,\overline{\rho_t^{\otimes 4}})=\dbraket{Q}{\overline{\rho_t^{\otimes 4}}}$.
The single-site Haar average over $\ket{\phi_i}$ yields
$\overline{(\ket{\phi_i}\!\bra{\phi_i})^{\otimes 4}} = \tfrac{1}{5}\Pi_{\mathrm{sym}}$,
where $\Pi_{\mathrm{sym}} = \tfrac{1}{4!}\sum_{\pi,\bs}\ket{\pi\cdot\bs}\!\bra{\bs}$ projects onto the replica-symmetric subspace; here $\bs\in\{0,1\}^{4}$ labels the computational basis of the four replicas, $r=1,\ldots,4$ is the replica index, and $\pi\in S_4$ acts by $(\pi\cdot\bs)_r:=s_{\pi(r)}$.
Diagrammatically,
\begin{equation} \label{eq:rho0}
\begin{aligned}
\dket{\overline{\rho_0^{\otimes 4}}} &\equiv \cdots\,\init\,\init\,\init\,\init\cdots, \\
\init &= \tfrac{1}{5}\dket{\Pi_{\mathrm{sym}}} = \frac{1}{5\cdot 4!}\sum_{\pi,\,\bs}\dket{\pi\cdot\bs,\bs}.
\end{aligned}
\end{equation}
Averaging a single two-qubit gate $U$ over the $\mathrm{U}(1)$-symmetric Haar measure defines the transfer-matrix building block~\cite{weingarten1978,collins2006},
\begin{equation} \label{eq:W4-main}
\mW \equiv \twositegate
\equiv \overline{U^{\otimes 4}\otimes U^{*\,\otimes 4}}.
\end{equation}
Each leg of $\mW$ carries the basis $\dket{\bs_{\ell},\ts_{\ell}}$ ($\ell=L,R$ labels left and right legs, respectively), with $\bs_{\ell},\ts_{\ell}\in\{0,1\}^{4}$ the ket and bra replica indices.
Because $U$ commutes with the charge operator, Haar invariance under $U \to U e^{\ii\theta\hat n_{\ell}}$, with $\hat n=(1-Z)/2$, imposes
\begin{align}
  & \dbra{\bs_L',\ts_L';\bs_R',\ts_R'}\mW (e^{\ii\theta\hat n_{\ell}})^{\otimes 4}\otimes(e^{-\ii\theta\hat n_{\ell}})^{\otimes 4}\dket{\bs_L,\ts_L;\bs_R,\ts_R} \nonumber \\
  &\quad=\; e^{\ii\theta(|\bs_{\ell}|-|\ts_{\ell}|)}\,\dbra{\bs_L',\ts_L';\bs_R',\ts_R'}\mW\dket{\bs_L,\ts_L;\bs_R,\ts_R},
\end{align}
where $|\bs_{\ell}|:=\sum_r s_{\ell,r}$ and $|\ts_{\ell}|:=\sum_r \tilde s_{\ell,r}$ are the total replica charges on leg $\ell$.
Since this must hold for all $\theta$, each leg obeys $|\bs_{\ell}|=|\ts_{\ell}|$, reducing the local dimension from $2^{8}=256$ to $\sum_{w=0}^{4}\binom{4}{w}^{2}=70$ (see the Supplemental Material for the complete selection rules~\cite{SM}).

Replacing each gate by $\mW$ yields the averaged state $\drho$ at depth $t$.
Individual circuit realizations break translation invariance, but the ensemble average restores it, so $\drho$ admits an infinite matrix-product-state (iMPS) representation with a two-site unit cell,
\begin{equation}
\drhoNetworkEquation \, ,
\label{eq:drho-network}
\end{equation}
in the Vidal canonical form~\cite{vidal2003,vidal2007}, with Schmidt matrices $\Lambda^{[i]}$ and site tensors $\Gamma^{[i]}$ ($i=0,1$).

In the thermodynamic limit, the relevant quantity is the entropy density $\ann{m_2}(t) = \ann{M_2}(t)/N$.
The operator $Q$ in Eq.~\eqref{eq:zeta} is a product of on-site operators,
\begin{equation}
\dbra{Q}
\equiv
\cdots\,\QQ\,\QQ\,\cdots, \quad
\QQ \equiv \tfrac{1}{4}\dbra{\sum_{q = I,X,Y,Z} q^{\otimes 4}}.
\end{equation}
The contraction with $Q$ gives a two-site transfer eigenvalue $\lambda_{\rm max;Q}$ [Fig.~\ref{fig:main-schematic}(c)], while the contraction with $I^{\otimes 4}$ gives the normalization eigenvalue $\lambda_{\rm max;tr}$ [Fig.~\ref{fig:main-schematic}(b)].
Thus the normalized averaged stabilizer purity is $\overline{\zeta_2(t)}
=
\left({\lambda_{\rm max;Q}}/{\lambda_{\rm max;tr}}\right)^{N/2}$,
and the entropy density reads
\begin{equation}
  \ann{m_2}(t) = -\frac{1}{2}\log_2 (\lambda_{\rm max;Q}/\lambda_{\rm max;tr}) - 1.
\end{equation}
Both the random initial state and the averaged gate $\mW$ are invariant under the $S_4$ permutation of the four replicas, so $\drho$ remains $S_4$-symmetric throughout. We exploit this non-Abelian symmetry to organize the iMPS into multiplets of $S_4$ irreducible representations~\cite{mcculloch2002,singh2010,singh2011,weichselbaum2012}. This block structure renders the compression step of the nonunitary iTEBD update [Fig.~\ref{fig:main-schematic}(d)] efficient; implementation details are collected in the End Matter and Supplemental Material~\cite{SM}.

\begin{figure*}[t]
  \centering
  \includegraphics[width=0.99\textwidth]{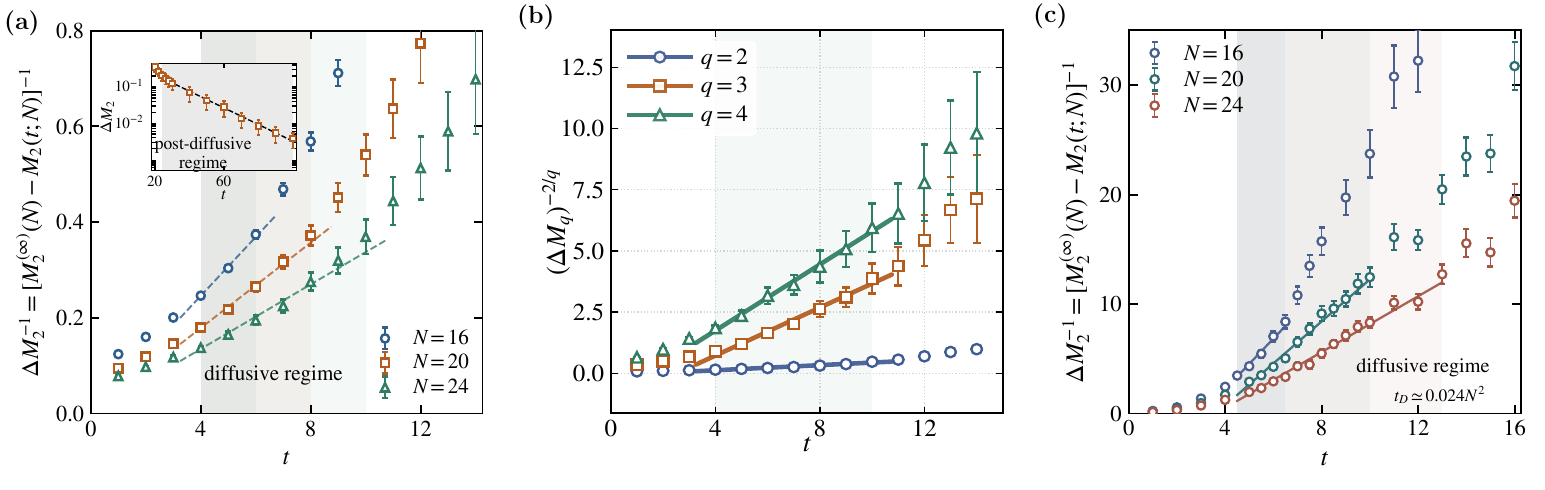}
  \caption{
  (a) $\mathrm{U}(1)$ random circuit: inverse gap $\Delta M_2^{-1}(t;N)$ for several $N$, linear in $t$ in the diffusive window. Inset: post-Thouless exponential tail on a semi-log scale.
  (b) Higher R\'enyi gaps in the same circuit at $N=24$, plotted as $[\Delta M_q]^{-2/q}$ ($q=2,3,4$) to linearize $\Delta M_q\propto t^{-q/2}$.
  (c) Mixed-field Ising chain [Eq.~\eqref{eq:mfim}]: same diffusive $t^{-1}$ scaling, with the diffusive window terminating at a Thouless-like crossover scale $t_D\simeq 0.024\,N^2$.
  Each point is averaged over 80 random realizations; error bars denote one standard error of the mean, and gray shading marks the fitting windows used for the diffusive guides.
  }
  \label{fig:diffusive-regimes}
\end{figure*}

{\it Diffusive universality class for nonstabilizerness.---}
Using the replica tensor network, we obtain the time evolution of $\ann{m_2}(t)$ [see Fig.~\ref{fig:main-schematic}(e); maximal total bond dimension $\chi_{\rm max}=800$, with the results converged for $\chi_{\rm max} \geq 600$].
At late times, $\ann{m_2}(t)$ approaches the thermodynamic-limit Haar value $m_2^{\rm Haar}=1$, and the gap satisfies
\begin{equation}
  \Delta m_2(t)\equiv 1-\ann{m_2}(t)\propto t^{-1}
  \label{eq:dm2-power}
\end{equation}
[see Fig.~\ref{fig:main-schematic}(e)]. This decay originates from the hydrodynamics of the conserved $\mathrm{U}(1)$ charge.
Splitting the on-site charge as $\langle Z_i\rangle_t=c^{\rm eq}_{Z_i}+u_i(t)$, with $\overline{c^{\rm eq}_{Z_i}}=0$ for our random product initial states, conservation makes the slow piece $u_i$ a diffusive hydrodynamic mode~\cite{rakovszky2018,nahum2018a}: $u_i(t)=\sum_j G(i-j,t)u_j(0)$ with $G(x,t)=(4\pi Dt)^{-1/2}e^{-x^2/(4Dt)}$ and diffusion constant $D$, giving
\begin{equation}
  \overline{u_i(t)^2}=C_0(8\pi Dt)^{-1/2}\propto t^{-1/2}
\end{equation}
for short-range initial correlations $\overline{u_i(0)u_j(0)}=C_0\,\delta_{ij}$.
The transverse operators $X_i,Y_i$ carry no slow component and relax exponentially~\cite{rakovszky2018,nahum2018a},
\begin{equation}
  \langle X_i\rangle_t-\langle X_i\rangle_\infty,\;\langle Y_i\rangle_t-\langle Y_i\rangle_\infty\sim e^{-t/\tau},\quad\tau=\mO(1).
  \label{eq:XY-decay}
\end{equation}
Pauli strings in the diagonal sector $\mathcal{P}_z\equiv\{I,Z\}^{\otimes N}$ relax slowly, whereas strings containing $X$ or $Y$ inherit the exponential decay of Eq.~\eqref{eq:XY-decay}. For $P\in\mathcal{P}_z\setminus\{I\}$, we approximate the slow relaxation as
\begin{equation}
\begin{aligned}
  \langle P\rangle_t&\simeq c_P^{\rm eq}+h_P(t),\\
  h_P(t)&\equiv\!\!\prod_{i:\,P_i=Z}\!\!u_i(t),\qquad P\in\mathcal{P}_z\setminus\{I\},
\end{aligned}
  \label{eq:Pauli-relaxation}
\end{equation}
where $c_P^{\rm eq}$ is the (Haar-like) equilibrium fluctuation. The factorization in Eq.~\eqref{eq:Pauli-relaxation} relies on the unitary circuit building up correlations only within a length scale $\mO(t)$, so it holds whenever the $Z$ insertions in $P$ are pairwise separated by more than $\mO(t)$. In the large-$N$ limit this is satisfied by the overwhelming majority of relevant Pauli strings: a string of finite weight $k$ has typical $Z$--$Z$ separation $N/k$, which greatly exceeds $t$ once $N\gg kt$, so almost all low-weight strings sit deep in the regime where Eq.~\eqref{eq:Pauli-relaxation} applies.

After the microscopic time $t\gg\tau$, the stabilizer purity decomposes as
\begin{align}
  \overline{\zeta_2(t)}
  &\approx \zeta_2^{\rm Haar}
  +4^{-N}\!\!\sum_{P\in\mathcal{P}_z\setminus\{I\}}\!\!\overline{6\,h_P(t)^2(c_P^{\rm eq})^2+h_P(t)^4},
  \label{eq:zeta-decompose}
\end{align}
where $\zeta_2^{\rm Haar}\equiv 4^{-N}\sum_P\overline{(c_P^{\rm eq})^4}$ is the equilibrium (Haar) value. The terms odd in the coarse-grained slow fields vanish for the leading, well-separated contributions because the initial charge fluctuations have zero mean and the late-time hydrodynamic fields are asymptotically Gaussian.
The Haar value satisfies $\zeta_2^{\rm Haar}=\mO(4^{-N})$, while $\overline{(c_P^{\rm eq})^2}=\mO(2^{-N})$ for $P\neq I$ and $h_P(t)=\mO(1)$ in $N$.
The cross term $\overline{h_P^2(c_P^{\rm eq})^2}$ is therefore smaller than $\overline{h_P^4}$ by a factor of $\mO(2^{-N})$ and can be dropped.
The dominant correction comes from $h_P^4$. Using $\sum_{P\in\mathcal{P}_z\setminus\{I\}}h_P^4=\prod_i(1+u_i^4)-1$ and expanding the late-time entropy density to leading order in $u$ gives
\begin{equation}
  \Delta m_2(t)=\frac{A}{N\ln 2}\sum_i\overline{u_i(t)^4}+\mO(u^8),\qquad A=\mO(1).
  \label{eq:dm2-u4}
\end{equation}
Since $u_i(t)$ is a sum of $\mO(\sqrt{Dt})$ independent initial-site contributions, the central limit theorem gives
\begin{equation}
  \overline{u_i(t)^4}=3\,\overline{u_i(t)^2}^{\,2}\propto t^{-1},
  \label{eq:u4-diff}
\end{equation}
which together with Eq.~\eqref{eq:dm2-u4} reproduces Eq.~\eqref{eq:dm2-power}.

{\it Diffusive and post-diffusive regimes.---}
The power-law decay [Eq.~\eqref{eq:dm2-power}] is established in the limit $N\to\infty$.
To see how it manifests at finite system size, we simulate $M_2(t;N)$ directly for finite chains using recently developed Pauli-string sampling methods~\cite{xiao2026sampling,huang2025b,sierant2026}, with random product initial states~\cite{SM}.
In contrast to the iTEBD calculation, which extracts the annealed quantity $\ann{M_2}=-\log_2\overline{\zeta_2}-N$, here we report the direct ensemble average $\overline{M_2(t;N)}$, and we have verified that the late-time scaling reported below is the same under either choice.
We define the gap relative to the measured long-time plateau,
\begin{equation}
  \Delta M_2(t;N)\equiv \overline{M_2^{(\infty)}(N)}-\overline{M_2(t;N)}.
\end{equation}
Unlike in the infinite system, $\Delta M_2$ shows two regimes [Fig.~\ref{fig:diffusive-regimes}(a)]:
\begin{equation}
  \Delta M_2(t;N)\propto
  \begin{cases}
    t^{-1}, & 2 \lesssim t\lesssim t_D, \\
    e^{-\gamma t}, & t\gg t_D.
  \end{cases}
\end{equation}
Furthermore, we find that the crossover time $t_D$ increases with $N$.
The crossover between the two regimes is naturally identified with the Thouless time.
The Fourier modes of the conserved density relax as $\tilde u_k(t)=e^{-Dk^2t}\tilde u_k(0)$.
The decay rate of the longest-wavelength mode $k_{\rm min}=2\pi/N$ gives the Thouless time $t_{D}\sim N^2/D$.
For $t<t_D$, the conserved modes are still diffusing, leading to the same scaling as in the infinite system [Eq.~\eqref{eq:dm2-power}].
After $t_D$, the conserved mode is nearly spatially uniform, so hydrodynamics no longer controls the gap and the decay becomes exponential, as also observed in nonstabilizerness dynamics without conserved modes~\cite{turkeshi2025}.

We then investigate higher stabilizer R\'enyi entropies within the diffusive window.
Similarly, we define the gap $\Delta M_q(t;N)\equiv \overline{M_q^{(\infty)}(N)}-\overline{M_q(t;N)}$.
We find that $\Delta M_q(t;N)\propto t^{-q/2}$ [Fig.~\ref{fig:diffusive-regimes}\,(b)].
This scaling can be understood by an argument similar to Eq.~\eqref{eq:dm2-u4}: the leading correction becomes $\overline{u_i(t)^{2q}}$, yielding the $t^{-q/2}$ scaling.
{\it Diffusive universality in a Hamiltonian system.---}
We now consider a nonintegrable Hamiltonian: the open-boundary mixed-field Ising chain
\begin{equation}
  H_{\rm MFIM}=\sum_{i=1}^{N} h_i Z_i + h_x \sum_{i=1}^{N}X_i + J\sum_{i=1}^{N-1}Z_i Z_{i+1},
  \label{eq:mfim}
\end{equation}
with $J=1$, $h_x=(\sqrt 5+5)/8$, and bulk longitudinal field $h_i=h_{\rm bulk}\equiv(\sqrt 5+1)/4$~\cite{kim2013}; the edge fields $h_1=h_N=h_{\rm bulk}-J$ suppress boundary effects.
Starting from a product initial state with a smoothly correlated charge profile~\cite{SM}, we evolve $|\psi_t\rangle=e^{-\ii H_{\rm MFIM}t}|\psi_0\rangle$ by the Chebyshev method~\cite{talezer1984} and compute the stabilizer R\'enyi entropy gap $\Delta M_2(t;N)$.
It follows the same algebraic decay $\Delta M_2(t;N)\propto t^{-1}$ as in the $\mathrm{U}(1)$ circuit [Fig.~\ref{fig:diffusive-regimes}\,(c)].
Because continuous-time dynamics allows us to resolve the crossover more finely, we further identify the upper edge of this regime as $t_D\simeq 0.024\,N^2$, quadratic in $N$ as expected from the diffusive Thouless picture.

The slow hydrodynamic mode is now the local energy density, which has nontrivial overlap with both $X_i$ and $Z_i$; Pauli strings supported on these operators inherit the same diffusive structure as the $Z$-strings of the $\mathrm{U}(1)$ case.
The hydrodynamic argument behind Eqs.~\eqref{eq:zeta-decompose}--\eqref{eq:u4-diff}, applied with energy in place of charge, then yields the same $t^{-1}$ scaling for $\Delta M_2(t;N)$ in the diffusive window.
The same diffusive logic has been identified in the entanglement growth and out-of-time-ordered correlators of chaotic systems with a slow conserved mode~\cite{khemani2018,rakovszky2018,nahum2018a}, and our result extends it to nonstabilizerness.
A previous finite-size study of both this model and the $\mathrm{U}(1)$ circuit reported larger and model-dependent power-law exponents~\cite{tirrito2025a}; this is naturally accounted for by single-power-law fits spanning the Thouless crossover, where the diffusive $t^{-1}$ window and the post-Thouless tail are mixed.

{\it Summary and outlook.---}
We have developed an $S_4$-adapted iTEBD method for computing the stabilizer R\'enyi entropy density of $\mathrm{U}(1)$-symmetric random circuits directly in the thermodynamic limit.
Using this method together with a hydrodynamic argument, we identify a diffusive universality class for the late-time approach of nonstabilizerness to its equilibrium value in chaotic many-body systems with a slow conserved density: the thermodynamic-limit entropy-density gap obeys $\Delta m_2\propto t^{-1}$, while finite-size R\'enyi gaps follow the hierarchy $\Delta M_q\propto t^{-q/2}$ within the diffusive window $\tau\ll t\ll t_D\sim N^2/D$.
Numerical simulations confirm this picture both in a $\mathrm{U}(1)$-symmetric random circuit and in an energy-conserving nonintegrable Ising chain.
Our work highlights the role of symmetry and hydrodynamics in quantum resource theory, complementing the established picture for entanglement entropy and operator spreading, and should be relevant for nonstabilizerness generation and approximate Haar-random state design in Hamiltonian dynamics~\cite{lami2025,zhang2025g,magni2025,varikuti2025}.

Several directions invite further study.
The early-time growth of stabilizer-entropy diagnostics starting from a stabilizer state has recently been connected to transport, with scaling as $t^{1/z}$ for dynamical exponent $z=1$, $3/2$, or $2$ depending on the underlying Hamiltonian~\cite{tirrito2025transport}; it would be interesting to characterize the corresponding late-time relaxation, in particular for different kinds of integrable models.
Whether other diagnostics of nonstabilizerness, such as robustness of magic, Bell magic, or entanglement-spectrum flatness~\cite{beverland2020,jiang2023d,haug2023c,dowling2025a,ahmadi2024,tirrito2024flatness}, share the diffusive universality identified here is also open.

{\it Note added.---}
After completion of this work, we became aware of a related study~\cite{liu2026pe}.
It also combines iMPS numerics with a hydrodynamic argument, but focuses on participation entropy rather than stabilizer R\'enyi entropy.

{\it Acknowledgments.---}
Z.X. thanks Sarang Gopalakrishnan, David A. Huse, Shuo Liu, and Kang Wang for helpful discussions.
The computations reported in this paper were performed using Princeton Research Computing resources.
Z.X. is supported by the Princeton Quantum Initiative Fellowship.

\bibliography{ref}

\onecolumngrid
\section*{End Matter}
\twocolumngrid

\subsection{Numerical techniques}
\label{sec:endmatter-numerics}
The local dimension $d_{\rm loc}=70$ per site makes a na\"ive iTEBD update costly.
We use $S_4$ symmetry adaptation to bring the computation under control.
Both the initial state and $\mW$ are invariant under $\dket{\bs,\ts}\mapsto\dket{\pi\cdot\bs,\pi\cdot\ts}$ for every $\pi\in S_4$, so $\drho$ remains $S_4$-symmetric throughout.
The on-site Hilbert space decomposes under $S_4$ as $\mathcal{H}_{\rm loc} \cong \bigoplus_{\lambda} \mathbb{C}^{m_\lambda}\otimes V_\lambda$, where $\lambda$ runs over the five irreducible representations (labeled by Young diagrams) of $S_4$ and the multiplicities are $(m_{[4]},m_{[1^4]},m_{[3,1]},m_{[2,1,1]},m_{[2,2]})=(9,1,11,5,6)$~\cite{SM}.
In a symmetry-adapted gauge~\cite{mcculloch2002,singh2010,singh2011,weichselbaum2012}, the bond spaces inherit the same block structure, and the on-site $\Gamma$ tensors obey the equivariance condition
\begin{equation}
\sum_{s'}(u_g)_{ss'}\,\Gamma^{[i],s'} = V_g^{[i\oplus1]}\,\Gamma^{[i],s}\,(V_g^{[i]})^{-1}, \qquad i=0,1,
\label{eq:equivariance-main}
\end{equation}
where the bond representations $V_g^{[i]}$ are linear (rather than projective), since the evolution starts from a product state and runs for finitely many steps~\cite{SM}.

The iTEBD update on each bond proceeds as illustrated in Fig.~\ref{fig:main-schematic}(d).
After applying $\mW$ to the two-site center tensor $\theta$ and recanonicalizing it (the update is nonunitary)~\cite{orus2008}, the recanonicalized tensor is split by singular value decomposition (SVD).
After fusing the physical and bond indices via Clebsch--Gordan coefficients, Schur's lemma guarantees that $\theta$ is block-diagonal in $\lambda$, so the decomposition reduces to independent SVDs on the smaller multiplicity blocks $\theta^{(\lambda)}$~\cite{SM}.
We retain $\chi_\lambda$ singular values per sector, giving the total bond dimension $\chi =\sum_\lambda \chi_\lambda\, d_\lambda$, with a cutoff $\chi\le\chi_{\rm max}$ on the total bond dimension and an additional discard threshold $10^{-8}$ on individual singular values.
The accuracy of the resulting iMPS is therefore controlled by $\chi_{\rm max}$.

\clearpage

\end{document}


\setcounter{equation}{0}
\setcounter{section}{0}
\setcounter{figure}{0}
\setcounter{table}{0}
\setcounter{page}{1}
\renewcommand{\theequation}{S\arabic{equation}}
\renewcommand{\thesection}{\Roman{section}}
\renewcommand{\thefigure}{S\arabic{figure}}
\renewcommand{\thetable}{S\arabic{table}}
\renewcommand{\bibnumfmt}[1]{[S#1]}
\renewcommand{\citenumfont}[1]{#1}

\title{Supplemental Material for \\ ``Diffusive Dynamics of Nonstabilizerness''}

\author{Zhenyu Xiao}
\email{zyxiao@princeton.edu}
\affiliation{Princeton Quantum Initiative, Princeton University, Princeton, New Jersey 08544, USA}

\author{Shinsei Ryu}
\affiliation{Department of Physics, Princeton University, Princeton, New Jersey 08544, USA}

\date{\today}

\maketitle

\tableofcontents

\section{Four-replica average of a single two-qubit gate}
\label{sec:single-gate}

\subsection{Haar average}
We evaluate the Haar average
\begin{equation}
\mW\;:=\;\overline{U^{\otimes 4}\otimes U^{*\,\otimes 4}},
\label{eq:defW4}
\end{equation}
which is a superoperator on the four-replica doubled Hilbert space, with $U$ on the ket space and $U^*$ on the bra space. 
The two-qubit Hilbert space decomposes into sectors of fixed total charge $Q=n_L+n_R\in\{0,1,2\}$,
\begin{equation}
\mathcal H_{LR}
=\mathrm{span}\!\left\{\ket{n_Ln_R}:n_L,n_R\in\{0,1\}\right\}
=\mathcal H_0\oplus\mathcal H_1\oplus\mathcal H_2,
\qquad
d_q:=\dim\mathcal H_q,\qquad(d_0,d_1,d_2)=(1,2,1),
\end{equation}
and a charge-conserving gate factorizes accordingly as
\begin{equation}
U=U_0\oplus U_1\oplus U_2,\qquad
U_0,\,U_2\sim\Haar\bigl(U(1)\bigr),\qquad
U_1\sim\Haar\bigl(U(2)\bigr).
\label{eq:gate-ensemble}
\end{equation}

Label the four replicas by $r=1,\dots,4$, and denote the ket and bra charge patterns by $\mathbf Q,\tilde{\mathbf Q}\in\{0,1,2\}^4$. Expanding each $U$ into charge blocks gives
\begin{equation}
\mW
=\sum_{\mathbf Q,\tilde{\mathbf Q}}\;
\overline{\,\bigotimes_{r=1}^{4}U_{Q_r}\otimes\bigotimes_{r=1}^{4}U_{\tilde Q_r}^{*}\,}.
\label{eq:W4-expand}
\end{equation}
Let $\mathcal P_{\mathbf Q}$ be the permutation that sorts the ket replicas by charge, placing all charge-$0$ replicas first, then charge-$1$, then charge-$2$, while preserving the original order within each sector. Define $\mathcal P_{\tilde{\mathbf Q}}$ analogously on the bra side, and set
\begin{equation}
\mathcal P_{\mathbf Q,\tilde{\mathbf Q}}
:=
\mathcal P_{\mathbf Q}\otimes\mathcal P_{\tilde{\mathbf Q}}.
\end{equation}
Then
\begin{equation}
\bigotimes_{r=1}^{4}U_{Q_r}
=
\mathcal P_{\mathbf Q}^{-1}
\left(
\bigotimes_{q=0}^{2}U_q^{\otimes n_q(\mathbf Q)}
\right)
\mathcal P_{\mathbf Q},
\qquad
\bigotimes_{r=1}^{4}U_{\tilde Q_r}^{*}
=
\mathcal P_{\tilde{\mathbf Q}}^{-1}
\left(
\bigotimes_{q=0}^{2}U_q^{*\,\otimes n_q(\tilde{\mathbf Q})}
\right)
\mathcal P_{\tilde{\mathbf Q}},
\label{eq:sorted-blocks}
\end{equation}
where $n_q(\mathbf Q)$ denotes the number of replicas in $\mathbf Q$ with charge
$q$, and
\begin{equation}
\bm n(\mathbf Q):=\bigl(n_0(\mathbf Q),n_1(\mathbf Q),n_2(\mathbf Q)\bigr),
\end{equation}
analogously for $\tilde{\mathbf Q}$. Substituting \eqref{eq:sorted-blocks}
into \eqref{eq:W4-expand} and using the independence of the blocks
$U_0,U_1,U_2$, the average factorizes sector by sector,
\begin{equation}
\overline{\,\bigotimes_{r=1}^{4}U_{Q_r}\otimes\bigotimes_{r=1}^{4}U_{\tilde Q_r}^{*}\,}
=\mathcal P_{\mathbf Q,\tilde{\mathbf Q}}^{-1}
\Biggl[\bigotimes_{q=0}^{2}
\overline{\,U_q^{\otimes n_q(\mathbf Q)}\otimes U_q^{*\,\otimes n_q(\tilde{\mathbf Q})}\,}\Biggr]
\mathcal P_{\mathbf Q,\tilde{\mathbf Q}}.
\label{eq:W4-split}
\end{equation}
For example, with $\mathbf Q=(1,0,1,2)$ the sort places replica $2$ first,
then replicas $1,3$, then replica $4$, so conjugation by $\mathcal P_{\mathbf
Q}$ sends $U_1\otimes U_0\otimes U_1\otimes U_2\mapsto U_0\otimes
U_1^{\otimes 2}\otimes U_2$, with $\bm n(\mathbf Q)=(1,2,1)$.

For each $q$, the Haar measure on $U_q$ is invariant under the global phase
shift $U_q\mapsto e^{\ii\phi_q}U_q$, under which the $q$-block in
\eqref{eq:W4-split} picks up the phase
$\exp\!\bigl\{\ii\phi_q\bigl[n_q(\mathbf Q)-n_q(\tilde{\mathbf Q})\bigr]\bigr\}$.
Averaging over $\phi_q$ therefore enforces charge balance
\begin{equation}
\bm n(\mathbf Q)=\bm n(\tilde{\mathbf Q}),
\label{eq:charge-balance}
\end{equation}
so only pairs $(\mathbf Q,\tilde{\mathbf Q})$ with matching charge multiplicities contribute.

For each such surviving pair, the sector-wise averages are given by the Weingarten calculus~\cite{weingarten1978,collins2006},
\begin{equation}
\overline{\,U_q^{\otimes n_q(\mathbf Q)}\otimes U_q^{*\,\otimes n_q(\mathbf Q)}\,}
=
\sum_{\sigma,\tau\in S_{n_q(\mathbf Q)}}
\Wg^{(d_q)}_{n_q(\mathbf Q)}\!\bigl(\sigma^{-1}\tau\bigr)\,
\dket{\sigma}_q\,\dbra{\tau}_q.
\label{eq:Wg-sector}
\end{equation}
In the one-dimensional sectors $q=0,2$, this formula reduces to the phase-balance constraint above and a scalar factor. When $n_1>d_1=2$, $\Wg^{(2)}_{n_1}$ is understood in the standard generalized (pseudoinverse) sense.
Here $\dket{\sigma}_q$ is the vectorized permutation state in the doubled $q$-sector: for any orthonormal basis $\{\ket{a}_q\}_{a=1}^{d_q}$ of $\mathcal H_q$,
\begin{equation}
\dket{\sigma}_q
:=
\sum_{\mathbf a\in\{1,\dots,d_q\}^{\,n_q(\mathbf Q)}}
\bigotimes_{r=1}^{n_q(\mathbf Q)}\ket{a_{\sigma(r)}}_q
\otimes
\bigotimes_{r=1}^{n_q(\mathbf Q)}\ket{a_r}_q.
\end{equation}
Substituting \eqref{eq:Wg-sector} into \eqref{eq:W4-split} and imposing charge balance \eqref{eq:charge-balance} gives
\begin{equation}
\mW
=
\sum_{\substack{\mathbf Q,\tilde{\mathbf Q}\\
\bm n(\mathbf Q)=\bm n(\tilde{\mathbf Q})}}
\mathcal P_{\mathbf Q,\tilde{\mathbf Q}}^{-1}
\Biggl[
\bigotimes_{q=0}^{2}
\sum_{\sigma_q,\tau_q\in S_{n_q(\mathbf Q)}}
\Wg^{(d_q)}_{n_q(\mathbf Q)}\!\bigl(\sigma_q^{-1}\tau_q\bigr)\,
\dket{\sigma_q}_q\,\dbra{\tau_q}_q
\Biggr]
\mathcal P_{\mathbf Q,\tilde{\mathbf Q}}.
\label{eq:W4-Wg}
\end{equation}
The sectors $q=0,2$ are one-dimensional, so each contributes a single $c$-number factor and only the charge-$1$ sector carries a nontrivial permutation structure. Write $n_1:=n_1(\mathbf Q)=n_1(\tilde{\mathbf Q})$ for the number of active charge-$1$ replicas. Suppressing the unique doubled states in the frozen sectors, we define the embedded charge-$1$ permutation states
\begin{equation}
\dket{\sigma;\mathbf Q,\tilde{\mathbf Q}}
:=
\mathcal P_{\mathbf Q,\tilde{\mathbf Q}}^{-1}\,\dket{\sigma}_{q=1},
\qquad
\dbra{\tau;\mathbf Q,\tilde{\mathbf Q}}
:=
\dbra{\tau}_{q=1}\,\mathcal P_{\mathbf Q,\tilde{\mathbf Q}}.
\label{eq:embedded-perm}
\end{equation}
Concretely, let $r_1<\dots<r_{n_1}$ be the active replica indices on the ket side (those with $Q_{r_j}=1$), and let $\tilde r_1<\dots<\tilde r_{n_1}$ denote the active replica indices on the bra side. Then $\dket{\sigma;\mathbf Q,\tilde{\mathbf Q}}$ is the vectorization of the partial permutation operator
\begin{equation}
V_{\sigma;\mathbf Q,\tilde{\mathbf Q}}
:=
\sum_{\mathbf a\in\{1,2\}^{\,n_1}}
\left(\bigotimes_{j=1}^{n_1}\ket{a_{\sigma(j)}}_{r_j,q=1}\right)
\left(\bigotimes_{j=1}^{n_1}\bra{a_j}_{\tilde r_j,q=1}\right),
\label{eq:V-sigma}
\end{equation}
which maps the active bra charge-$1$ sites to the active ket charge-$1$ sites, permuting them by $\sigma$. Substituting into \eqref{eq:W4-Wg} and keeping only the nontrivial charge-$1$ sum, we obtain
\begin{equation}
\mW
\;=\;
\sum_{\substack{
\bm n(\mathbf Q)=\bm n(\tilde{\mathbf Q})\\
\sigma,\tau\in S_{n_1(\mathbf Q)}
}}
\Wg^{(2)}_{n_1(\mathbf Q)}\!\bigl(\sigma^{-1}\tau\bigr)\,
\dket{\sigma;\mathbf Q,\tilde{\mathbf Q}}\,
\dbra{\tau;\mathbf Q,\tilde{\mathbf Q}},
\label{eq:W4-final}
\end{equation}
where the combined sum runs over $(\mathbf Q,\tilde{\mathbf Q})$ satisfying the
charge-balance condition \eqref{eq:charge-balance} and over
$\sigma,\tau\in S_{n_1(\mathbf Q)}$.
Here $\sigma$ and $\tau$ permute only the ordered list of active charge-$1$ replicas within the fixed $(\mathbf Q,\tilde{\mathbf Q})$ block, and all nontrivial structure of $\mW$ resides in this block.

\subsection{Basis expansion}
\label{sec:basis-expansion}
We now write \eqref{eq:W4-final} as explicit matrix elements in the doubled
computational basis, which is the form used by the matrix-product-operator (MPO) implementation.
Throughout, unprimed/primed labels denote input/output, and tildes
distinguish the bra copy from the ket copy.

A basis element of the two-site four-replica doubled space is the
vectorization of $\ket{\bs}\!\bra{\ts}$,
\begin{equation}
\dket{\bs,\ts}
:=
\ket{\bs}\otimes\ket{\ts},
\qquad
\bs=(s_{\ell,r})_{\ell\in\{L,R\},\,r=1,\dots,4}\in\{0,1\}^{2\times 4},
\end{equation}
with $\ts$ defined identically; $\ell\in\{L,R\}$ labels the two physical
qubits and $r=1,\dots,4$ the replicas. The superoperator $\mW$ is a
$2^{16}\times 2^{16}$ matrix, or equivalently a four-leg tensor with each
leg of dimension $2^8=256$. We now show that $\mathrm{U}(1)$ symmetry
reduces each leg to dimension $70$ and derive the explicit form of
$\dbra{\bs',\ts'}\mW\dket{\bs,\ts}$.

\paragraph*{Reduction to $70$ dimensions per leg.}
The $\mathrm{U}(1)$-conserved Haar measure is invariant under
right-multiplication $U\mapsto Ug(\theta)$ with
$g(\theta)=e^{\ii\theta\hat n_L}$, under which
\begin{equation}
\dbra{\bs',\ts'}\,
U^{\otimes 4}g(\theta)^{\otimes 4}\otimes
U^{*\,\otimes 4}g(-\theta)^{\otimes 4}\,
\dket{\bs,\ts}
=
e^{\ii\theta\,(|\bs_L|-|\ts_L|)}\,
\dbra{\bs',\ts'}\,U^{\otimes 4}\otimes U^{*\,\otimes 4}\,\dket{\bs,\ts},
\label{eq:right-phase}
\end{equation}
where $|\bs_\ell|:=\sum_{r=1}^{4}s_{\ell,r}$ and
$|\ts_\ell|:=\sum_{r=1}^{4}\tilde s_{\ell,r}$ are the total occupation
numbers on qubit $\ell$ summed over the four replicas. Averaging over $U$
and using Haar invariance,
\begin{equation}
\dbra{\bs',\ts'}\mW\dket{\bs,\ts}
=
e^{\ii\theta\,(|\bs_L|-|\ts_L|)}\,
\dbra{\bs',\ts'}\mW\dket{\bs,\ts}
\qquad\forall\,\theta,
\end{equation}
so $\mW$ vanishes unless $|\bs_L|=|\ts_L|$. Repeating with
$g(\theta)=e^{\ii\theta\hat n_R}$ gives $|\bs_R|=|\ts_R|$, and
left-multiplication invariance $U\mapsto g(\theta)U$ yields
$|\bs'_\ell|=|\ts'_\ell|$ for $\ell\in\{L,R\}$, where
$|\bs'_\ell|:=\sum_{r=1}^{4}s'_{\ell,r}$ and
$|\ts'_\ell|:=\sum_{r=1}^{4}\tilde s'_{\ell,r}$. In total, $\mW$
vanishes unless $|\bs_\ell|=|\ts_\ell|$ on each of its four legs, reducing
each to
\begin{equation}
d_{\mathrm{loc}}
=\sum_{w=0}^{4}\binom{4}{w}^{2}=70
\label{eq:70d}
\end{equation}
basis elements. Thus, $\mW$ is a four-leg tensor of dimension $70$ per leg.

\paragraph*{Ket--bra charge balance.}
The charge-balance condition $\bm n(\mathbf Q)=\bm n(\tilde{\mathbf Q})$
derived in \eqref{eq:charge-balance} further constrains the basis: for each
pair charge $q\in\{0,1,2\}$, the number of replicas carrying that charge
must match between the ket and bra copies. In the computational basis this
reads
\begin{equation}
\sum_{r=1}^{4}\delta_{s_{L,r}+s_{R,r},\,q}
=
\sum_{r=1}^{4}\delta_{\tilde s_{L,r}+\tilde s_{R,r},\,q},
\qquad q=0,1,2,
\label{eq:charge-balance-basis}
\end{equation}
and the same condition must hold on the output side with
$(\bs,\ts)\to(\bs',\ts')$. $\mW$ vanishes unless both the input and output
basis elements satisfy \eqref{eq:charge-balance-basis}.

\paragraph*{Input--output charge conservation.}
For each replica $r$, define the pair charges
\begin{equation} \label{eq:Qr}
Q_r:=s_{L,r}+s_{R,r},\quad
Q'_r:=s'_{L,r}+s'_{R,r},\quad
\tilde Q_r:=\tilde s_{L,r}+\tilde s_{R,r},\quad
\tilde Q'_r:=\tilde s'_{L,r}+\tilde s'_{R,r}
\;\in\{0,1,2\}.
\end{equation}
Since the original gate commutes with the total charge,
$[U,\,e^{\ii\theta(\hat n_L+\hat n_R)}]=0$, this symmetry can be applied
on each replica independently. Conjugating by
$e^{\ii\theta(\hat n_{L,r}+\hat n_{R,r})}$ on replica $r$ alone gives
\begin{equation}
\dbra{\bs',\ts'}\mW\dket{\bs,\ts}
=
e^{\ii\theta\,(Q_r-Q'_r)}\,
\dbra{\bs',\ts'}\mW\dket{\bs,\ts}
\qquad\forall\,\theta,
\end{equation}
so $Q_r=Q'_r$ for every $r$. The same argument on the bra side gives
$\tilde Q_r=\tilde Q'_r$, hence
\begin{equation} \label{eq:Qr-final}
Q_r=Q'_r,\qquad\tilde Q_r=\tilde Q'_r,\qquad r=1,\dots,4.
\end{equation}

\paragraph*{Permutation symmetry.}
Another symmetry is the invariance under simultaneous permutation of the four
replicas on both ket and bra sides. For any $\pi\in S_4$, define
\begin{equation}
\ket{\pi\cdot\bs,\pi\cdot\ts}
:=
\bigotimes_{r=1}^{4}
\ket{s_{L,\pi(r)},s_{R,\pi(r)}}\otimes
\ket{\tilde s_{L,\pi(r)},\tilde s_{R,\pi(r)}},
\end{equation}
and similarly for $\dbra{\pi\cdot\bs',\pi\cdot\ts'}$. Then,
\begin{equation}
\dbra{\pi\cdot\bs',\pi\cdot\ts'}\mW\dket{\pi\cdot\bs,\pi\cdot\ts}
=
\dbra{\bs',\ts'}\mW\dket{\bs,\ts}.
\label{eq:replica-perm}
\end{equation}

We now derive the nonzero matrix elements of $\mW$ on the subspace
singled out by the constraints \eqref{eq:charge-balance-basis},
\eqref{eq:Qr-final}. Since the output charge patterns are
fixed to the input ones, it suffices to specify the ket and bra charge
patterns $Q_r$ and $\tilde Q_r$ on the input side. Within any such pattern
the replicas with $Q_r\in\{0,2\}$ are frozen: the charge alone fixes the bit
pair to $00$ or $11$ on both input and output, leaving no dynamical content.
All nontrivial structure therefore resides on the charge-$1$ replicas.

Let $r_1<\dots<r_{n_1}$ denote the ket-side replicas with $Q_r=1$, and $\tilde r_1<\dots<\tilde r_{n_1}$ the corresponding
bra-side replicas with $\tilde Q_r=1$; the two lists share the common
length $n_1$ by the charge-balance condition
\eqref{eq:charge-balance-basis}. 
They also label the output-side
charge-$1$ replicas since $Q_r=Q'_r$ and $\tilde Q_r=\tilde Q'_r$. 
On such
a replica the bit pair is either $01$ or $10$, and the state is fully
specified by the value on the left qubit (the right one being fixed by
charge conservation). Reading off the left-qubit value at the ordered
charge-$1$ positions,
\begin{equation}
x_j:=s_{L,r_j},\quad
x'_j:=s'_{L,r_j},\quad
\tilde x_j:=\tilde s_{L,\tilde r_j},\quad
\tilde x'_j:=\tilde s'_{L,\tilde r_j},
\qquad j=1,\dots,n_1,
\end{equation}
and collecting them in order, we obtain four length-$n_1$ bitstrings
\begin{equation}
\mathbf x:=(x_1,\dots,x_{n_1}),\qquad
\mathbf x':=(x'_1,\dots,x'_{n_1}),\qquad
\tilde{\mathbf x}:=(\tilde x_1,\dots,\tilde x_{n_1}),\qquad
\tilde{\mathbf x}':=(\tilde x'_1,\dots,\tilde x'_{n_1}),
\end{equation}
that encode all active degrees of freedom; the right-qubit values are
determined by $s_{R,r_j}=1-x_j$, and analogously for the primed and tilded
versions.

In this parametrization, \eqref{eq:W4-final} gives the nonzero matrix elements
\begin{equation}
\dbra{\bs',\ts'}\mW\dket{\bs,\ts}
=
\begin{cases}
1, & n_1=0,\\[1mm]
\displaystyle
\sum_{\sigma,\tau\in S_{n_1}}
\Wg^{(2)}_{n_1}\!\bigl(\sigma^{-1}\tau\bigr)\,
\delta_{\mathbf x',\,\sigma\cdot\tilde{\mathbf x}'}\,
\delta_{\mathbf x,\,\tau\cdot\tilde{\mathbf x}},
& n_1\ge 1,
\end{cases}
\label{eq:matrix-final}
\end{equation}
where $S_{n_1}$ acts on length-$n_1$ strings by
$(\sigma\cdot\mathbf y)_j:=y_{\sigma(j)}$. The two Kronecker deltas say that,
after ordering the charge-$1$ replicas, the ket strings $\mathbf x,\mathbf x'$
must match the bra strings $\tilde{\mathbf x},\tilde{\mathbf x}'$ up to
permutations $\sigma,\tau\in S_{n_1}$, each weighted by the Weingarten function
$\Wg^{(2)}_{n_1}(\sigma^{-1}\tau)$.

\section{Representation-theoretic structure of the local \texorpdfstring{$S_4$}{S4} action}
\label{sec:s4-structure}

The tensor $\mW$ is invariant under simultaneous permutation of the four replica labels on the ket and bra sides [Eq.~\eqref{eq:replica-perm}], and the initial state shares this symmetry, so the averaged state $\dket{\overline{\rho_t^{\otimes 4}}}$ remains $S_4$-invariant at all times.
In this section, we collect the representation-theoretic data of $S_4$ needed for the symmetry-adapted infinite time-evolving block decimation (iTEBD) algorithm described in Sec.~\ref{sec:algorithm} (see, e.g., Ref.~\cite{serre1977}).

\subsection{Irreducible representations of \texorpdfstring{$S_4$}{S4}}

The irreducible representations (irreps) of $S_4$ are labeled by the partitions of $4$,
\begin{equation}
\operatorname{Irr}(S_4) = \{[4],\,[1^4],\,[3,1],\,[2,1,1],\,[2,2]\},
\end{equation}
with dimensions $d_\lambda = 1,1,3,3,2$ respectively.
The character table is given in Table~\ref{tab:s4-character}.

\begin{table}[t]
\caption{\label{tab:s4-character} Character table of $S_4$. Rows are the irreducible representations of $S_4$, labeled by integer partitions $\lambda$ of $4$ (equivalently, Young diagrams); columns are the conjugacy classes, labeled by the cycle type of a representative permutation.}
\begin{center}
\begin{tabular}{lccccc}
\hline\hline
$\lambda$ & $(1^4)$ & $(2,1,1)$ & $(2,2)$ & $(3,1)$ & $(4)$ \\
\hline
$[4]$ & $1$ & $1$ & $1$ & $1$ & $1$ \\
$[1^4]$ & $1$ & $-1$ & $1$ & $1$ & $-1$ \\
$[3,1]$ & $3$ & $1$ & $-1$ & $0$ & $-1$ \\
$[2,1,1]$ & $3$ & $-1$ & $-1$ & $0$ & $1$ \\
$[2,2]$ & $2$ & $0$ & $2$ & $-1$ & $0$ \\
\hline\hline
\end{tabular}
\end{center}
\end{table}

We write $D_\lambda(g)$ for the representation matrices.
The adjacent transpositions
$s_{12}=(12)$, $s_{23}=(23)$, $s_{34}=(34)$ generate $S_4$, so each irrep is
fixed by the three matrices $D_\lambda(s_{12})$, $D_\lambda(s_{23})$,
$D_\lambda(s_{34})$. The one-dimensional irreps have
$D_{[4]}(s_{ij})=1$ and $D_{[1^4]}(s_{ij})=-1$. Since
$[2,1,1]\cong[1^4]\otimes[3,1]$, we have
$D_{[2,1,1]}(s_{ij})=-D_{[3,1]}(s_{ij})$, so only the $[3,1]$ and $[2,2]$
matrices need to be specified independently. In the orthogonal Young basis
(right-action convention $D_\lambda(\pi\sigma)=D_\lambda(\sigma)D_\lambda(\pi)$),
\begin{equation}
D_{[3,1]}(s_{12})=
\begin{pmatrix}
\tfrac12 & \tfrac{\sqrt3}{2} & 0 \\[2pt]
\tfrac{\sqrt3}{2} & -\tfrac12 & 0 \\[2pt]
0 & 0 & 1
\end{pmatrix},
\quad
D_{[3,1]}(s_{23})=
\begin{pmatrix}
1 & 0 & 0 \\[2pt]
0 & \tfrac13 & \tfrac{2\sqrt2}{3} \\[2pt]
0 & \tfrac{2\sqrt2}{3} & -\tfrac13
\end{pmatrix},
\quad
D_{[3,1]}(s_{34})=
\begin{pmatrix}
\tfrac12 & -\tfrac{1}{2\sqrt3} & -\sqrt{\tfrac23} \\[2pt]
-\tfrac{1}{2\sqrt3} & \tfrac56 & -\tfrac{\sqrt2}{3} \\[2pt]
-\sqrt{\tfrac23} & -\tfrac{\sqrt2}{3} & -\tfrac13
\end{pmatrix},
\end{equation}
\begin{equation}
D_{[2,2]}(s_{12})=
\begin{pmatrix}
\tfrac12 & -\tfrac{\sqrt3}{2} \\[2pt]
-\tfrac{\sqrt3}{2} & -\tfrac12
\end{pmatrix},
\qquad
D_{[2,2]}(s_{23})=
\begin{pmatrix}
\tfrac12 & \tfrac{\sqrt3}{2} \\[2pt]
\tfrac{\sqrt3}{2} & -\tfrac12
\end{pmatrix},
\qquad
D_{[2,2]}(s_{34})=D_{[2,2]}(s_{12}).
\end{equation}

\subsection{Tensor products and Clebsch--Gordan coefficients}

For irreps $V_\alpha$ and $V_\beta$, the tensor product decomposes as
\begin{equation}
V_{\alpha}\otimes V_{\beta}
\cong
\bigoplus_{\lambda\in\operatorname{Irr}(S_4)}
\mathbb{C}^{N_{\alpha\beta}^{\lambda}}\otimes V_{\lambda},
\qquad
N_{\alpha\beta}^{\lambda}
=
\frac{1}{24}
\sum_{C \subset S_4}
|C|\,\chi_{\alpha}(C)\,\chi_{\beta}(C)\,\chi_{\lambda}(C).
\label{eq:tp-decomp}
\end{equation}
The Clebsch--Gordan (CG) coefficients are the matrix elements of the unitary basis change from the uncoupled basis $|i,j\rangle=|i\rangle_\alpha\otimes|j\rangle_\beta$ to the coupled basis $|\lambda,\mu;a\rangle$,
\begin{equation}
|\lambda,\mu;a\rangle
=
\sum_{i,j}
(C_{\alpha\beta})^{\lambda,\mu;a}_{ij}\,|i,j\rangle,
\label{eq:cg-def}
\end{equation}
where $\lambda$ labels the irrep, $\mu=1,\dots,N_{\alpha\beta}^{\lambda}$ the multiplicity copy, and $a=1,\dots,d_\lambda$ a component within $V_\lambda$.
In the coupled basis, the group action becomes block-diagonal,
\begin{equation}
C_{\alpha\beta}\,
\bigl(D_{\alpha}(g)\otimes D_{\beta}(g)\bigr)
=
\left[
\bigoplus_{\lambda\in\operatorname{Irr}(S_4)}
\left(\mathbb{I}_{N_{\alpha\beta}^{\lambda}}\otimes D_{\lambda}(g)\right)
\right]
C_{\alpha\beta},
\qquad g\in S_4,
\label{eq:CG-equivariance}
\end{equation}
with $\mu$ a spectator index on which $g$ does not act.

To construct the coupled basis, one introduces the matrix units
\begin{equation}
E_{ab}^{(\lambda)}
=
\frac{d_\lambda}{24}
\sum_{g\in S_4}
[D_\lambda(g^{-1})]_{ab}\,
\bigl(D_\alpha(g)\otimes D_\beta(g)\bigr).
\label{eq:matrix-units}
\end{equation}
The operator $E_{11}^{(\lambda)}$ projects onto a subspace of dimension $N_{\alpha\beta}^{\lambda}$; choosing an orthonormal basis $\{u_\mu^{(\lambda)}\}$ of its image, the coupled basis vectors are $|\lambda,\mu;a\rangle\propto E_{a1}^{(\lambda)}\,u_\mu^{(\lambda)}$.
In practice, we obtain $C_{\alpha\beta}$ numerically by simultaneously block-diagonalizing the three generator matrices $D_\alpha(s_{ij})\otimes D_\beta(s_{ij})$ for $s_{ij}\in\{s_{12},s_{23},s_{34}\}$.

\subsection{Decomposition of the local \texorpdfstring{$70$}{70}-dimensional physical space}

The local physical basis is the neutral four-replica doubled basis
\begin{equation}
\mathcal{B}_{\mathrm{loc}}
=
\left\{
(\mathbf{k},\mathbf{b})\in\{0,1\}^{4}\times\{0,1\}^{4}:
\sum_{r=1}^{4} k_r = \sum_{r=1}^{4} b_r
\right\},
\label{eq:local-basis}
\end{equation}
whose $70$ elements satisfy the neutrality constraint derived in \eqref{eq:70d}.
Replica permutations act by simultaneously relabeling both copies,
\begin{equation}
R_{\mathrm{loc}}(\pi)\,|(\mathbf{k},\mathbf{b})\rangle
=
|(\pi\cdot\mathbf{k},\pi\cdot\mathbf{b})\rangle,
\qquad \pi\in S_4,
\label{eq:S4-action}
\end{equation}
defining a $70$-dimensional representation of $S_4$.
Its isotypic decomposition is
\begin{equation}
\mathcal{H}_{\mathrm{local}}
\cong
\bigoplus_{\lambda\in\operatorname{Irr}(S_4)}
\mathbb{C}^{m_\lambda}\otimes V_{\lambda},
\qquad
(m_{[4]},\,m_{[1^4]},\,m_{[3,1]},\,m_{[2,1,1]},\,m_{[2,2]})
=(9,\,1,\,11,\,5,\,6),
\label{eq:schur-decomp}
\end{equation}
where $V_\lambda$ carries the irrep $\lambda$, while $\mathbb C^{m_\lambda}$ is the multiplicity space on which $S_4$ acts trivially.
The decomposition is verified by the dimension check $9\!\cdot\!1+1\!\cdot\!1+11\!\cdot\!3+5\!\cdot\!3+6\!\cdot\!2=70$.
This block structure organizes the local physical leg into $S_4$ sectors and underlies the symmetry-adapted truncation scheme used in the iTEBD algorithm.

\section{Symmetry-preserving nonunitary TEBD}
\label{sec:algorithm}

The averaged state is represented as an infinite matrix product state (iMPS) in the Vidal canonical form~\cite{vidal2003,vidal2007} with a two-site unit cell.
We use the iTEBD algorithm~\cite{vidal2007}, extended to nonunitary gates following Ref.~\cite{orus2008}.
The $S_4$ symmetry is exploited using the framework of symmetry-adapted tensor networks~\cite{singh2010,singh2011}.
Each update consists of three steps: (i)~application of a two-site gate to the local center tensor, (ii)~exact recanonicalization of the resulting two-site block, and (iii)~bond compression within $S_4$ symmetry sectors. Only step~(iii) introduces truncation.

\subsection{Symmetry-adapted Vidal form}

We use the Vidal form throughout. For a two-site unit cell,
\begin{equation}
\dket{\Psi}
=
\sum_{\{s_n\}}
\Bigl(
\cdots\,
\Lambda^{[0]}\Gamma^{[1],s_{-1}}\Lambda^{[1]}\Gamma^{[0],s_0}
\Lambda^{[0]}\Gamma^{[1],s_1}\Lambda^{[1]}\Gamma^{[0],s_2}\Lambda^{[0]}
\,\cdots
\Bigr)
|\cdots s_{-1}s_0s_1s_2\cdots\rangle.
\label{eq:period-two}
\end{equation}
Here $\Gamma^{[i],s}$ carries one local physical index and two virtual indices,
while $\Lambda^{[i]}$ is the diagonal Schmidt matrix on the bond between sites
$i$ and $i\oplus1$. The canonical conditions are
\begin{equation}
\sum_{s_i}
\bigl(\Lambda^{[i\oplus1]}\Gamma^{[i],s_i}\bigr)^\dagger
\bigl(\Lambda^{[i\oplus1]}\Gamma^{[i],s_i}\bigr)
= \mathbb{I},
\qquad
\sum_{s_i}
\bigl(\Gamma^{[i],s_i}\Lambda^{[i]}\bigr)
\bigl(\Gamma^{[i],s_i}\Lambda^{[i]}\bigr)^\dagger
= \mathbb{I},
\qquad (i=0,1).
\label{eq:canonical}
\end{equation}
In addition, $\Tr[(\Lambda^{[i]})^2]=1$, so that $(\Lambda^{[i]}_{\alpha\alpha})^2$ are the Schmidt weights on the bipartition $(-\infty,i]\,|\,[i+1,\infty)$.

The initial state and the two-site gate are $S_4$ invariant, so the iMPS can be kept $S_4$ invariant after every update:
\begin{equation}
u_g^{\otimes\mathbb Z}\dket{\Psi}=\dket{\Psi},
\qquad g\in S_4.
\end{equation}
This invariance is implemented directly on the MPS legs. The physical leg is
the $70$-dimensional neutral one-site replica space of
Sec.~\ref{sec:s4-structure}. In the $S_4$-adapted basis,
\begin{equation}
u_g
=
\bigoplus_{\lambda\in\operatorname{Irr}(S_4)}
\bigl(I_{m_\lambda}\otimes D_\lambda(g)\bigr),
\qquad
\mathcal H_{\mathrm{local}}
\cong
\bigoplus_{\lambda\in\operatorname{Irr}(S_4)}
\left(\mathbb C^{m_\lambda}\otimes V_\lambda\right),
\label{eq:physical-rep}
\end{equation}
where the multiplicities $m_\lambda$ are given in
\eqref{eq:schur-decomp}. Thus a physical MPS index has the form
$s=(\lambda_s,\mu_s,a_s)$, where $\lambda_s$ is the irrep sector,
$\mu_s=1,\dots,m_{\lambda_s}$ labels the multiplicity copy, and
$a_s=1,\dots,d_{\lambda_s}$ labels the irrep component.

In a symmetry-adapted gauge, the same $S_4$ action is carried by the virtual
bond spaces. Let $V_g^{[i]}$ denote the representation on the bond between
sites $i$ and $i\oplus1$. The $S_4$ action on the physical leg of each Vidal tensor is equivalent to conjugation by representations $V_g^{[i]}$ on its two virtual legs,
\begin{equation}
\sum_{s'}(u_g)_{ss'}\,\Gamma^{[i],s'}
=
V_g^{[i\oplus1]}\,\Gamma^{[i],s}\,\bigl(V_g^{[i]}\bigr)^{-1},
\qquad i=0,1.
\label{eq:equivariance}
\end{equation}
For the states considered here these virtual actions can be chosen as linear
representations,
\begin{equation}
V_g^{[i]}\,V_h^{[i]} = V_{gh}^{[i]},
\qquad i=0,1,
\label{eq:virtual-rep}
\end{equation}
so each bond space also decomposes into $S_4$ sectors,
\begin{equation}
V_g^{[i]}
=
\bigoplus_{\lambda\in\operatorname{Irr}(S_4)}
\bigl(I_{\chi_\lambda^{[i]}}\otimes D_\lambda(g)\bigr),
\qquad i=0,1,
\label{eq:bond-rep}
\end{equation}
where $\chi_\lambda^{[i]}$ is the bond multiplicity in sector $\lambda$.
A bond index therefore has the same representation-theoretic structure as a
physical index,
$\alpha=(\lambda_\alpha,\nu_\alpha,a_\alpha)$, with
$\nu_\alpha=1,\dots,\chi_{\lambda_\alpha}^{[i]}$ labeling the bond
multiplicity copy.
In the numerical implementation, each bond index carries such a three-part label $(\lambda,\nu,a)$, so that all tensor contractions and truncations can be performed block by block within each irrep sector.

\subsection{Gate application and blocked recanonicalization}

Without loss of generality, consider the update on the bond $\Lambda^{[0]}$ between sites $0$ and $1$. From Eq.~\eqref{eq:period-two}, the Vidal center tensor on this bond has components
\begin{equation}
\theta_{\alpha s_0 s_1 \gamma}
=
\sum_{\beta}
\Lambda^{[1]}_{\alpha\alpha}\,
\Gamma^{[0],s_0}_{\alpha\beta}\,
\Lambda^{[0]}_{\beta\beta}\,
\Gamma^{[1],s_1}_{\beta\gamma}\,
\Lambda^{[1]}_{\gamma\gamma}.
\label{eq:theta-components}
\end{equation}
This tensor is the exact two-site wavefunction extracted from the canonical
iMPS. It carries an $S_4$ representation on its physical legs and another on its virtual legs, with the two actions related by Eq.~\eqref{eq:equivariance}.

If the previous step used a reduced local basis, $\theta$ is first embedded
back into the full local space. The gate $\mW$ [Eq.~\eqref{eq:W4-final}] is a $70^2\times 70^2$ matrix on the pair of full local indices $(p_0,p_1)$, and its application gives
\begin{equation}
(\theta'_{\mathrm{full}})_{\alpha p_0 p_1 \gamma}
= \sum_{s_0,s_1} \mW_{p_0 p_1,\,s_0 s_1}\,\theta_{\alpha s_0 s_1 \gamma}.
\label{eq:gate-apply}
\end{equation}
Because $\mW$ is not unitary, the resulting tensor is no longer in canonical form and must be recanonicalized~\cite{orus2008}.

To recanonicalize, we treat $(p_0,p_1)$ as a single blocked physical index and remove the outer Schmidt matrix,
\begin{equation}
C_{\alpha p_0 p_1 \gamma}
=
\sum_{\alpha'}
((\Lambda^{[1]})^{-1})_{\alpha\alpha'}\,
(\theta'_{\mathrm{full}})_{\alpha' p_0 p_1 \gamma},
\label{eq:blocked-C}
\end{equation}
where the inverse is on the support of $\Lambda^{[1]}$.
The blocked transfer maps are
\begin{align}
(\mathcal{E}_R(X))_{\alpha\alpha'}
&=
\sum_{p_0,p_1,\gamma,\gamma'}
C_{\alpha p_0 p_1 \gamma}\,
X_{\gamma\gamma'}\,
C^*_{\alpha' p_0 p_1 \gamma'},
\nonumber\\
(\mathcal{E}_L(X))_{\gamma\gamma'}
&=
\sum_{p_0,p_1,\alpha,\alpha'}
C^*_{\alpha p_0 p_1 \gamma}\,
X_{\alpha\alpha'}\,
C_{\alpha' p_0 p_1 \gamma'} .
\label{eq:transfer}
\end{align}
Their dominant fixed points satisfy $\mathcal{E}_R(r)=\kappa\,r$ and $\mathcal{E}_L(l)=\kappa\,l$.
Since each virtual leg of $C$ transforms under a direct sum of $S_4$ irreps [Eq.~\eqref{eq:bond-rep}], the fixed points $l$ and $r$ must also respect this block structure: they are block-diagonal in the irrep label $\lambda$ and act only on the multiplicity indices within each sector.

After rescaling $C\leftarrow\kappa^{-1/2}C$, factorize
$l=X_L^\dagger X_L$ and $r=X_R X_R^\dagger$. The canonical gauge is obtained
from the overlap singular value decomposition (SVD)
\begin{equation}
X_L\,X_R = U\,\tilde\Lambda\,V^\dagger,
\label{eq:overlap-svd}
\end{equation}
where $\tilde\Lambda$ is the recanonicalized Schmidt matrix on the outer bond.
Like $\Lambda^{[1]}$, it is block-diagonal in the irrep label: $\tilde\Lambda=\bigoplus_\lambda \tilde\Lambda^{(\lambda)}\otimes I_{d_\lambda}$, and each bond index retains the label $(\lambda,\nu,a)$.
The canonical blocked center tensor is then
\begin{equation}
(\theta_{\mathrm{can}})_{\alpha p_0 p_1 \gamma}
=
\sum_{\alpha',\gamma'}
(U^\dagger X_L)_{\alpha\alpha'}\,
C_{\alpha' p_0 p_1 \gamma'}\,
(X_R V)_{\gamma'\gamma},
\label{eq:theta-can}
\end{equation}
which is exact before truncation. Equivalently, the corresponding blocked
Vidal tensor is
\[
(\Gamma_{\mathrm{can}})_{\alpha p_0 p_1 \gamma}
=
(\tilde\Lambda_\alpha)^{-1}
(\theta_{\mathrm{can}})_{\alpha p_0 p_1 \gamma}
(\tilde\Lambda_\gamma)^{-1},
\]
with the inverses taken on the support of $\tilde\Lambda$.

\subsection{Bond compression}

The symmetry-preserving truncation follows the non-Abelian framework of Refs.~\cite{mcculloch2002,weichselbaum2012}, adapted to our nonunitary setting.
To split the canonical blocked tensor back into a two-site MPS, first regard it as a
matrix from the right fused space to the left fused space. The two fused spaces
are
\begin{equation}
\mathcal K_0=\mathcal V^{[1]}\otimes{\mathcal H}_{\mathrm{local}},
\qquad
\mathcal K_1={\mathcal H}_{\mathrm{local}}\otimes\mathcal V^{[1]},
\end{equation}
where $\mathcal H_{\mathrm{local}}$ is the full $70$-dimensional local space and $\mathcal V^{[1]}$ is the outer bond space. Their
Clebsch--Gordan decompositions are
\begin{equation}
\mathcal K_i
\cong
\bigoplus_{\lambda\in\operatorname{Irr}(S_4)}
\left(\mathbb C^{n_{i,\lambda}}\otimes V_\lambda\right),
\qquad i=0,1.
\label{eq:fused-spaces}
\end{equation}
Here $n_{i,\lambda}$ counts the allowed multiplicity channels after coupling
the outer-bond irrep with the physical irrep. For example,
\begin{equation}
n_{0,\lambda}
=
\sum_{\lambda_\alpha,\lambda_0}
\chi_{\lambda_\alpha}^{[1]}\,
m_{\lambda_0}\,
N_{\lambda_\alpha\lambda_0}^{\lambda},
\end{equation}
with the analogous expression for $n_{1,\lambda}$.

Let $C^0$ and $C^1$ be the unitary CG basis changes from the uncoupled indices
$(\alpha,p_0)$ and $(p_1,\gamma)$ to coupled indices
$(\lambda,\eta,a)$, where $\eta$ labels the multiplicity channel and
$a=1,\ldots,d_\lambda$ labels the irrep component. In this coupled basis,
\begin{align}
\Theta_{(\lambda,\eta_0,a),(\lambda',\eta_1,a')}
&=
\sum_{\alpha,p_0,p_1,\gamma}
(C^0_{\lambda,\eta_0,a;\alpha p_0})^*\,
(\theta_{\mathrm{can}})_{\alpha p_0 p_1\gamma}\,
C^1_{\lambda',\eta_1,a';p_1\gamma}.
\label{eq:theta-coupled}
\end{align}
Because the $S_4$ actions on the two fused spaces of $\theta_{\mathrm{can}}$ are both representations, Schur's lemma gives
\begin{equation}
\Theta_{(\lambda,\eta_0,a),(\lambda',\eta_1,a')}
=
\delta_{\lambda\lambda'}\delta_{aa'}\,
(\theta^{(\lambda)})_{\eta_0,\eta_1}.
\label{eq:theta-schur}
\end{equation}
Thus the SVD of the full fused matrix reduces exactly to independent SVDs on
the multiplicity matrices:
\begin{equation}
\Theta
=
\bigoplus_{\lambda\in\operatorname{Irr}(S_4)}
\theta^{(\lambda)}\otimes I_{d_\lambda}.
\label{eq:theta-block}
\end{equation}
The block SVDs are
\begin{equation}
(\theta^{(\lambda)})_{\eta_0,\eta_1}
=
\sum_{\nu=1}^{\min(n_{0,\lambda},n_{1,\lambda})}
U^{(\lambda)}_{\eta_0,\nu}\,
S^{(\lambda)}_{\nu}\,
(V^{(\lambda)}_{\eta_1,\nu})^*,
\label{eq:bond-svd}
\end{equation}
and the largest $\tilde\chi_\lambda$ singular values are retained in each total
irrep sector $\lambda$. Each retained multiplicity singular vector carries the
whole irrep space $V_\lambda$; equivalently, the corresponding singular value
appears with degeneracy $d_\lambda$ in the full uncoupled matrix. The new bond
space is
\begin{equation}
\mathcal V'
=
\bigoplus_{\lambda\in\operatorname{Irr}(S_4)}
\left(\mathbb C^{\tilde\chi_\lambda}\otimes V_\lambda\right).
\end{equation}
The new Schmidt matrix on the inner bond is
\begin{equation}
\Lambda'=\bigoplus_{\lambda\in\operatorname{Irr}(S_4)} S^{(\lambda)}\otimes I_{d_\lambda},
\end{equation}
after the sector-wise truncation.
With the convention of Eq.~\eqref{eq:cg-def}, the inverse transformation from the
coupled basis back to the uncoupled fused indices uses $C^0$ on the left fused
space and $(C^1)^*$ on the right fused space. The updated single-site tensors
are therefore recovered as
\begin{equation}
(\Gamma^{[0],p_0}_{\mathrm{new}})_{\alpha\beta}
=
(\tilde\Lambda_\alpha)^{-1}
\sum_{\eta_0}
C^0_{\lambda_\beta,\eta_0,a_\beta;\alpha p_0}\,
U^{(\lambda_\beta)}_{\eta_0,\nu_\beta},
\qquad
(\Gamma^{[1],p_1}_{\mathrm{new}})_{\beta\gamma}
=
\sum_{\eta_1}
(V^{(\lambda_\beta)}_{\eta_1,\nu_\beta})^*\,
(C^1_{\lambda_\beta,\eta_1,a_\beta;p_1\gamma})^*\,
(\tilde\Lambda_\gamma)^{-1},
\label{eq:gamma-update}
\end{equation}
where the new inner bond index $\beta=(\lambda_\beta,\nu_\beta,a_\beta)$ runs over the retained singular vectors, with $\nu_\beta=1,\dots,\tilde\chi_{\lambda_\beta}$ and $a_\beta=1,\dots,d_{\lambda_\beta}$.
In the numerical implementation, the irrep label $\lambda$ is recorded for every new bond index, so that subsequent contractions and truncations remain block-diagonal in the irrep sectors.

The recanonicalization step [Eqs.~\eqref{eq:blocked-C}--\eqref{eq:theta-can}]
is exact; approximation enters only through the bond SVD truncation. This truncation discards multiplicity states while retaining the full irrep component $a=1,\dots,d_\lambda$, so $S_4$ symmetry is preserved exactly at every step.

Concretely, for the bond SVD, we retain $\chi_\lambda$ singular values per sector, giving total bond dimension $\chi=\sum_\lambda \chi_\lambda d_\lambda$, with a hard cutoff $\chi\le\chi_{\rm max}$ and an additional discard threshold $10^{-8}$ on individual singular values. The accuracy of the resulting iMPS is therefore controlled by $\chi_{\rm max}$.

\section{Numerical details for finite-size simulations}

To compute the stabilizer R\'enyi entropy at finite size, we use the fast Walsh--Hadamard transform (FWHT) Pauli sampler of Ref.~\cite{xiao2026sampling}. Pauli operators are partitioned into $2^N$ families $\{P_{x,z}\}_{z\in\mathbb{F}_2^N}$ labeled by $x\in\mathbb{F}_2^N$, and a single FWHT produces all $\{\langle\psi|P_{x,z}|\psi\rangle\}_z$ at cost $\mO(N\,2^N)$. We then Monte Carlo sample $\mathcal{N}\ll 2^N$ values of $x$ and accumulate the partial moments $m_{\alpha;x}=\sum_z|\langle\psi|P_{x,z}|\psi\rangle|^{2\alpha}/d^{\alpha}$, giving an unbiased estimator of $\zeta_\alpha=\sum_P |c_P|^{2\alpha}/d^{\alpha}$. The reported $M_\alpha$ is obtained from the plug-in estimator $M_\alpha=(1-\alpha)^{-1}\log_2\zeta_\alpha-N$. The $x=0$ family is always included exactly because it contains the identity, whose contribution is anomalously large. For $N<20$ we sweep $x$ over the entire $\mathbb{F}_2^N$ and obtain $M_\alpha$ exactly; for $N\ge 20$ we use $\mathcal{N}=10^4$ Monte Carlo samples per realization.

For both the $\mathrm{U}(1)$ random circuit and the mixed-field Ising chain, the initial states are random product states of the form
\begin{equation}
  |\psi_0\rangle
  = \bigotimes_{i=1}^{N}
  \left[
  \sqrt{\tfrac12+\eta_i}\,|0\rangle_i
  + e^{\ii\varphi}\sqrt{\tfrac12-\eta_i}\,|1\rangle_i
  \right],
\end{equation}
with $\varphi=0$ for the $\mathrm{U}(1)$ circuit and $\varphi=\pi/2$ for the Ising chain, so that $\langle Z_i\rangle_0=2\eta_i$. The classical field $\eta_i$ is drawn, for each realization, from a zero-mean Gaussian process with smooth, short-range correlations,
\begin{equation}
  \overline{\eta_i\eta_j} = \sigma^2 \exp\!\left[-\frac{(i-j)^2}{2\xi^2}\right],
\end{equation}
with $(\sigma,\xi)=(0.20, 4)$ for the $\mathrm{U}(1)$ circuit and $(\sigma,\xi)=(0.25, 5)$ for the Ising chain. The Gaussian profile is clipped to the interval $[-1/2,1/2]$. Each realization is a product state, so connected quantum correlators vanish on the initial state.

\bibliography{ref}